\documentclass[twocolumn,english,aps,showpacs,superscriptaddress]{revtex4-1}
\usepackage[latin9]{inputenc}
\usepackage{verbatim}
\usepackage{amsmath}
\usepackage{amssymb}
\usepackage{graphicx}
\usepackage{esint}

\makeatletter

\newcommand*\LyXThinSpace{\,\hspace{0pt}}

\usepackage{bm}
\usepackage{xcolor}
\usepackage{subfigure}
\usepackage{array}

\providecommand{\tabularnewline}{\\}

\@ifundefined{textcolor}{}{%
 \definecolor{BLACK}{gray}{0}
 \definecolor{WHITE}{gray}{1}
 \definecolor{RED}{rgb}{1,0,0}
 \definecolor{GREEN}{rgb}{0,1,0}
 \definecolor{BLUE}{rgb}{0,0,1}
 \definecolor{CYAN}{cmyk}{1,0,0,0}
 \definecolor{MAGENTA}{cmyk}{0,1,0,0}
 \definecolor{YELLOW}{cmyk}{0,0,1,0}
}

\usepackage{bm}

\newcommand{\beq}{\begin{equation}}
\newcommand{\eeq}{\end{equation}}
\newcommand{\bea}{\begin{eqnarray}}
\newcommand{\eea}{\end{eqnarray}}

\newcommand{\fvec}[1]{\boldsymbol{#1}}

\newcommand{\rmd}{{\rm d}}

\newcolumntype{x}[1]{>{\centering\arraybackslash}p{#1}}

\usepackage{babel}

\makeatother

\usepackage{babel}
\begin{document}

\title{Superconductivity in FeSe: the role of nematic order}

\author{Jian Kang}

\selectlanguage{english}%

\affiliation{National High Magnetic Field Laboratory, Florida State University,
Tallahassee, Florida 32304 USA}

\author{Rafael M. Fernandes}

\affiliation{School of Physics and Astronomy, University of Minnesota, Minneapolis,
MN 55455, USA}

\author{Andrey Chubukov}

\affiliation{School of Physics and Astronomy, University of Minnesota, Minneapolis,
MN 55455, USA}
\begin{abstract}
Bulk FeSe is a special iron-based material in which superconductivity
emerges inside a well-developed nematic phase. We present a microscopic
model for this nematic superconducting state, which takes into account
the mixing between $s-$wave and $d-$wave pairing channels and the
changes in the orbital spectral weight promoted by the sign-changing
nematic order parameter. We show that nematicity
only weakly affects $T_c$, but gives rise to  $\cos2\theta$
variation of the pairing gap on the hole pocket,
 whose magnitude and size agrees
with ARPES and  STM data.
 We further show that nematicity increases the weight of $d_{xz}$ orbital on the hole pocket, and increases (reduces) the weight of $d_{xy}$ orbital on $Y$ ($X$) electron pocket.
\end{abstract}
\maketitle
\emph{Introduction.} Superconductivity in FeSe has attracted a lot
of attention recently because this material holds the promise to reveal
new physics not seen in other Fe-based superconductors~\cite{recent}.
The pairing in FeSe emerges at $T\leq8$ K from a state with a well-defined
nematic order, which develops at a much higher
 $T_{s}
\sim 90$ K.
 Because nematic order breaks the $C_{4}$ tetragonal symmetry down
to $C_{2}$, it mixes the $s-$wave and $d-$wave pairing channels~\cite{Fernandes13,Kang14,livanas}.
As a result, the pairing gap on the $\Gamma-$centered hole pocket,
$\Delta(\theta)$, has both $s-$wave and $d-$wave components, $\Delta(\theta)=\Delta_{1}+\Delta_{2}\cos{2\theta}$,
where $\Delta_{1}$ and $\Delta_{2}$ are $C_{4}-$symmetric functions
of $\cos{4\theta}$.
 This gap form is generic, but
 the relative
sign between $\Delta_{1}$ and $\Delta_{2}$
depends on details of the pairing interaction and the structure of the nematic
order.

\begin{figure}[htbp]
\includegraphics[width=0.6\columnwidth]{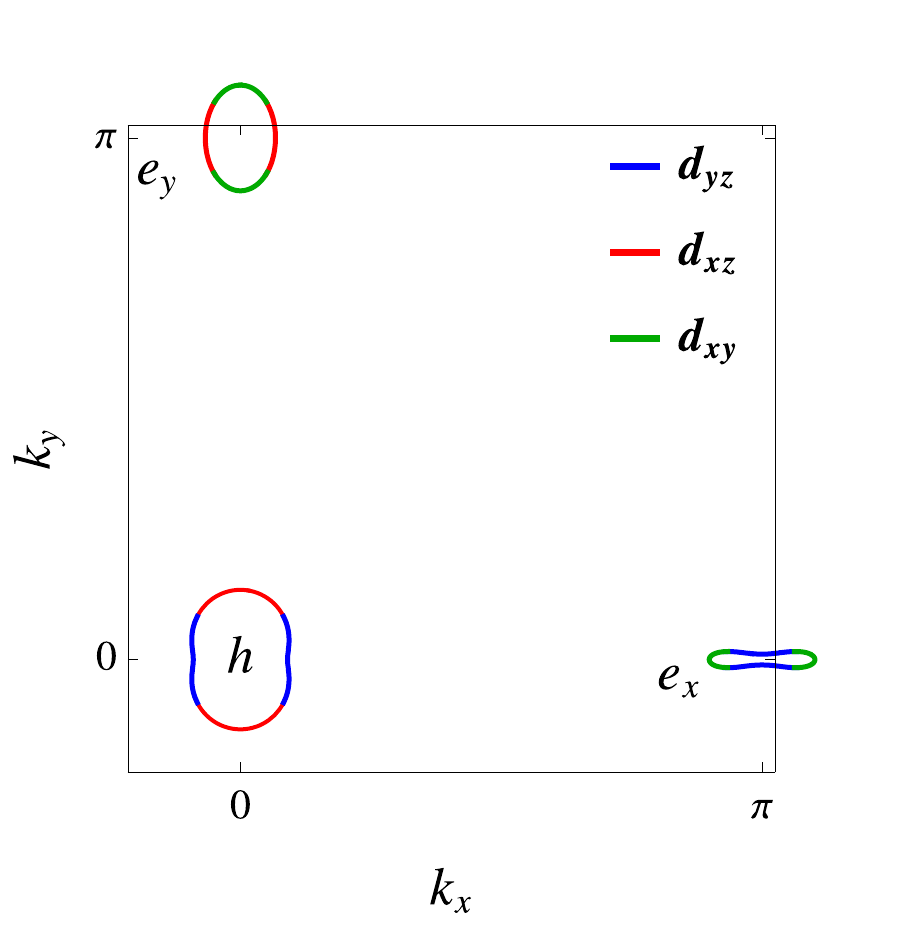}
\caption{The Fermi surface and its orbital content in the nematic phase of
FeSe. In the 1-Fe Brillouin zone there is a hole ($h$) pocket centered
at $\Gamma$/$Z=(0,0)$ and two electron pockets $X$ and $Y$ centered
at $(\pi,0)$ and $(0,\pi)$, respectively. STM and ARPES data \cite{Dong16,BrianSTM17}
show that the $h$ pocket is an ellipse elongated along $Y$,
and that the $X$ electron pocket has a peanut-type form with the
minor axis along the $Y$ direction. }
\label{Fig:FS}
\end{figure}

The $\cos{2\theta}$ gap anisotropy on the hole pocket (``$h$\char`\"{}
pocket in Fig.~\ref{Fig:FS}) has been probed recently by angle resolved
photoemission spectroscopy (ARPES)~\cite{Dong16,shin,Borisenko18,Zhou18,Rhodes18}
and scanning
tunneling microscopy (STM)~\cite{BrianSTM17,Wirth17}
measurements.
 These probes have shown
  that
(i) The gap is larger along the direction towards the $X$ electron pocket
made out of $d_{yz}$ and $d_{xy}$ orbitals, than towards the $Y$
pocket made out of $d_{xz}$ and $d_{xy}$ orbitals (Fig.~\ref{Fig:FS}); and (ii)
The magnitude of the gap on the $X$ pocket correlates with the
 weight of
 the  $d_{yz}$ orbital component.
  This led to the proposal~\cite{BrianSTM17}
that the pairing glue in FeSe is orbital-selective and predominantly
involves fermions from the $d_{yz}$ orbital.

To support this argument, Refs.~\cite{BrianSTM17,Brian17} analyzed
the pairing problem within BCS theory, using the static interaction
in the spin channel as the glue. They argued that the
observed
 gap anisotropy can  be reproduced only if one phenomenologically re-calibrates
the interactions
 on
  $d_{xz}$, $d_{yz}$, and $d_{xy}$
orbitals and set the interaction on the $d_{yz}$ orbital to be the
strongest. This was done by introducing
 phenomenologically
 different
 constant
 $Z-$factors for each orbital. A constant $Z$ does not give rise to incoherence,
but affects the magnitudes of the interactions on different
orbitals.

In this paper we reconsider this issue. We argue that one has to distinguish
 the difference between
  $d_{xy}$ orbital
and $d_{xz}/d_{yz}$ orbitals, and
 the difference
between
$d_{xz}$ and $d_{yz}$ orbitals. The dressed interactions
on the $d_{xy}$ orbital and on the $d_{xz}/d_{yz}$ orbitals are
not equal already in the tetragonal phase and
flow to different
values as one progressively integrates out contributions from high-energy
fermions~\cite{RG,ckf,ruiqi}. Adding different
 constant $Z$-factors to
the $d_{xy}$ and $d_{xz}/d_{yz}$ orbitals is
 a legitimate
way to incorporate these high-energy renormalizations into the low-energy
model. On the other hand, the interactions on
$d_{xz}$ and $d_{yz}$
orbitals become different only in the presence of nematic order.
 The latter is
 of order 10 meV (Refs. \cite{amalia,borisenko,BrianSTM17}),
  much smaller
than the electronic bandwidth.
As a result, the $d_{xz}/d_{yz}$ splitting
is a low-energy phenomenon which
 should be fully captured within the low-energy model, without introducing
 phenomenologically $Z_{xz}\neq Z_{yz}$.

In our approach we depart from the tetragonal phase with the
 $\Gamma$/$Z$-,
$X$-, and $Y$-centered Fermi pockets in the 1-Fe Brillouin zone.
We use the low-energy model of Ref.~\cite{Vafek13} to parametrize
the dispersion near these three points, and the model of Ref.~\cite{ckf}
for the $d_{xz}/d_{yz}$ pairing interactions in the $s-$wave and
$d-$wave channels. We introduce a two-component $d-$wave nematic
order parameter ${\bar \Phi} = (n_{d_{xz}}-n_{d_{yz}})/2=({\bar \Phi}_{h},{\bar \Phi}_{e})$,
where $h$ and $e$ refer to hole and electron pockets. It reconstructs
the dispersion and the Fermi pockets to the ones shown in Fig.~\ref{Fig:FS}.
 ARPES and STM data~\cite{Dong16,shin,Borisenko18,Zhou18,Rhodes18,BrianSTM17,amalia,borisenko,Shibauchi15}
 reveal
 an ellipsoidal
hole pocket elongated along the $Y$ direction, and a peanut-like
$X$ electron pocket. A simple analysis shows that such Fermi surfaces
emerge if $\Phi_{h}>0$ and $\Phi_{e}<0$, i.e.~the nematic order
changes sign between hole and electron pockets. This sign change is
consistent with
theoretical analysis~\cite{ckf,kontani,Fanfarillo16,Benfatto18}. We take as an input
 the results of earlier studies~\cite{RG,ckf,ruiqi,kontani,scalapino,chubukov_12}
that the largest pairing interaction at low-energies is
between hole and electron pockets. This interaction
 is angle-dependent in the band basis and has $s^{+-}$
and $d_{x^2-y^2}$
 components $U_{s}$ and $U_{d}$, respectively.
  $U_s$ is larger, and in the absence of nematicity the system develops
  $s^{+-}$ order, which changes sign between hole and electron pockets.
We dress
up $U_{s}$ and $U_{d}$ by coherence factors associated with the
nematic order,
 solve
  the gap equation, and
obtain $T_{c}$ and the structure of the superconducting gap~\cite{comm_a}.

Our results show that $T_{c}$ is only moderately affected by nematicity,
 but the nematic order gives rise to a sizable anisotropy of the gap on both hole
  and electron pockets.
 This is consistent with:
 (i) The phase diagram
of S-doped $\mathrm{FeSe_{1-x}S_{x}}$, which shows that $T_{c}$
changes little around
$x<0.17$, when nematic order disappears;
 and (ii) Thermal conductivity,
specific heat, and STM
 data~\cite{Taillefer16,Meingast17,Matsuda,hanaguri},
which show that the gap anisotropy changes drastically between $x<0.17$
and $x>0.17$.

For the gap on the hole pocket we find
$\Delta(\theta_{h})\approx\Delta_{h}(1 + \alpha\cos{2\theta_{h}}+\beta\cos{4\theta_{h}})$,
where the $\cos{2\theta_{h}}$ term is induced by nematicity.
To leading order in
${\bar \Phi}$,  $\alpha
 \propto
 \left(4|\Phi_{e}| -(U_{d}/U_{s})\Phi_{h}\right)$, where
  $\Phi_{h,e}$ are dimensionless orbital orders, normalized to the corresponding Fermi energies
  (see~\cite{SM} and Eq (\ref{7}) below).
The $\Phi_h$ term reflects the nematicity-induced mixing between the $s$ and $d$ pairing components whereas the
 $\Phi_e$ term is related to the nematicity-induced redistribution of orbital weight on the electron pockets.
The experimental angular dependence
of the gap is reproduced when the $\Phi_{e}$ term is larger. We computed
 $\Phi_{h,e}$ using band structure parameters
 which fit the ARPES data for the $Z$ pocket ($k_z = \pi$)~\cite{SM}
and found
  $|\Phi_{e}|\sim 0.1,~\Phi_{h}\sim 0.3$.
 Combining this with the fact that $U_s \geq U_d$,
  we see that
  $\alpha$ is positive.
 A
positive $\alpha$ can be interpreted as if nematicity makes the pairing
interaction between the $\Gamma$ and $X$ pockets stronger than between
the $\Gamma$ and $Y$ pockets. We emphasize that this effect is fully
captured within the low-energy model.

\begin{figure}[htbp]
\centering
\includegraphics[width=0.8\columnwidth]{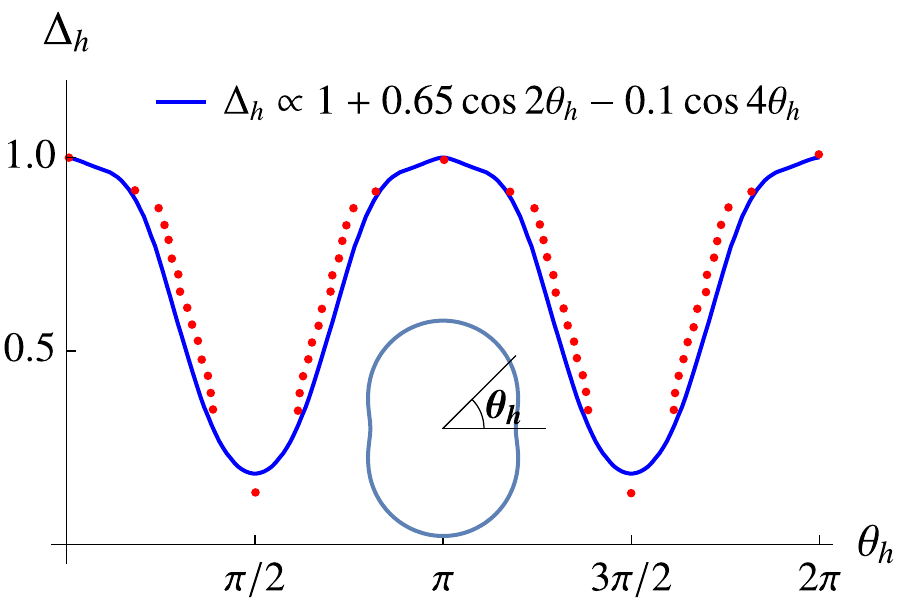} \caption{
Angular dependence of the pairing gap on the hole pocket obtained
by numerically solving the gap equations with band structure parameters
and nematic order parameters fitted to ARPES data above $T_{c}$.
The gap maximum is along the $\Gamma-X$ direction, consistent with
STM and ARPES data~\cite{Dong16,Shibauchi15,shin,BrianSTM17,Borisenko18,Zhou18,Rhodes18}. Points are STM data from
Ref. \cite{BrianSTM17}. The gap function
$\Delta (\theta_h) = \Delta_h (1 + \alpha \cos{2\theta_h} + \beta \cos{4\theta_h})$  has the
$\cos{2\theta}$ terms induced by nematicity (which we explicitly computed),  as well as $C_{4}$-symmetric
anisotropic $\cos{4\theta}$ terms present already in the tetragonal
phase due to spin-orbit coupling~\cite{Vafek13}
and/or due to dressing of
the pairing interaction by high-energy fermions~\cite{scalapino,ckf}.}

\label{Fig:HoleGap}
\end{figure}

In Fig. \ref{Fig:HoleGap} we show
the
 calculated $\Delta (\theta_h)$
 along with the gap anisotropy extracted from the STM data\cite{BrianSTM17}.We see that the agreement is quite good. We found equally good agreement with the ARPES data for the Z-pocket~\cite{Dong16,shin,Borisenko18,Zhou18}.
 Whether STM is probing the $Z$ ($k_z =\pi$) or the $\Gamma$ pocket ($k_z =0$) is difficult to determine, since STM data is likely
  averaged over $k_z$. We also computed the gap at the smaller $\Gamma$ pocket ($k_z=0$) and found a smaller gap with a weaker anisotropy. This arises because the dimensionless $\Phi_h$ is larger for smaller pockets and because, unlike the $Z$ pocket, the whole $\Gamma$ pocket has predominantly $d_{xz}$ character~\cite{SM}. A smaller gap at $\Gamma$  agrees with the ARPES data in~\cite{Dong16,Borisenko18} but not with ~\cite{Rhodes18}.

\emph{Low-energy model.} We consider a quasi-2D model of bulk FeSe,
which in the tetragonal phase has two
corrugated  cylindrical
hole pockets, centered at the
$k_{x,y}=0$ with the largest cross-section at $k_z=\pi$ and the smallest at $k_z =0$ (Refs. \cite{Dong16,amalia,Watson15,Borisenko18,Rhodes18})
  and two
cylindrical
electron pockets centered at $(\pi,0, k_z)$ and $(0,\pi, k_z)$ in the Fe-only Brillouin zone
($X$ and
$Y$ pockets).   The hole pockets are made out
primarily of $d_{xz}$ and $d_{yz}$ orbitals, the $X$ pocket is
made primarily out of $d_{yz}$ and $d_{xy}$ orbitals, and the $Y$
pocket, of $d_{xz}$ and $d_{xy}$ orbitals. We model the low-energy
electronic structure on each pocket by spinors, following Ref. \cite{Vafek13,review_fc}.
We choose parameters such that the larger hole pocket $h$ has $d_{xz}$
character along the $Y$ direction and $d_{yz}$ character
along the $X$ direction, consistent with ARPES experiments~\cite{Dong16,shin,amalia,Watson15,Borisenko18,Rhodes18,Shibauchi15}

The band operators for $h$, $X$, and $Y$ pockets are expressed
in terms of the orbital operators as
\begin{align}
h & =d_{yz}\cos{\phi_{h}}+d_{xz}\sin{\phi_{h}}\nonumber \\
e_{X} & =-id_{yz}\cos{\phi_{X}}+d_{xy}\sin{\phi_{X}}\nonumber \\
e_{Y} & =id_{xz}\cos{\phi_{Y}}+d_{xy}\sin{\phi_{Y}}\ ,\label{2}
\end{align}
In the tetragonal phase, the $h$-pocket is nearly circular and in the absence of spin-orbit coupling (SOC) $\phi_{h}\approx\theta_{h}$,
where $\theta_{h}$ is the angle measured with respect to the $X$
axis. On electron pockets, to a good approximation $\cos{\phi_{X,Y}}=A\sin{\theta_{X,Y}}$,
$\sin{\phi_{X,Y}}=(1-A^{2}\sin^{2}{\theta_{X,Y}})^{1/2}$, where $A<1$
and $\theta_{X}$ ($\theta_{Y}$) is the angle measured with respect
to the $X$ ($Y$) direction~\cite{ruiqi,Kang14}.

In the nematic phase we introduce momentum-dependent $d-$wave nematic
order
with components $\pm {\bar \Phi}_{h}$  (plus sign on $d_{xz}$ orbital)
and
${\bar \Phi}(Y)=-{\bar \Phi}(X)={\bar \Phi}_{e}$. For simplicity we neglect the $d_{xy}$
component of the nematic order~\cite{vafek_fernandes,ruiqi}. Eqs.~(\ref{2})
still hold in the presence of nematicity, but the relations
between $\phi_{h},\phi_{X},\phi_{Y}$ and the angles along the Fermi
surfaces become different and are obtained by the diagonalization
of the corresponding quadratic Hamiltonians. For the hole pocket,
we define the dimensionless $\Phi_{h}$ via $\cot{2\phi_{h}}=\cot{2\theta_{h}}-2\Phi_{h}/\sin{2\theta_{h}}$,
 again
in the absence of SOC (the full expressions including SOC are presented in~\cite{SM}). Roughly, $\Phi_h ={\bar \Phi}_h/E_F$. For the same ${\bar \Phi}_h$,
$\Phi_h$ is larger on the $\Gamma$ pocket than on $Z$ pocket, because $E_F$ is smaller
 in the former.

\begin{figure}[htbp]
\centering
\subfigure[\label{Fig:Orbital:Hole}]{\includegraphics[width=0.47\columnwidth]{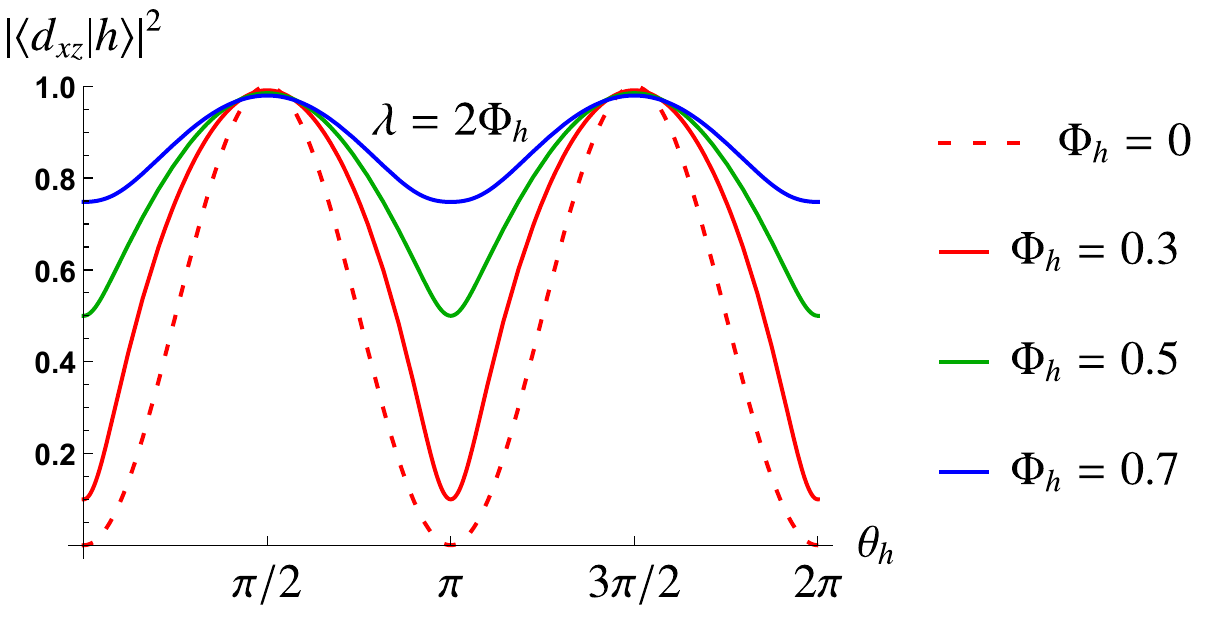}}
\subfigure[\label{Fig:Orbital:Ele}]{\includegraphics[width=0.47\columnwidth]{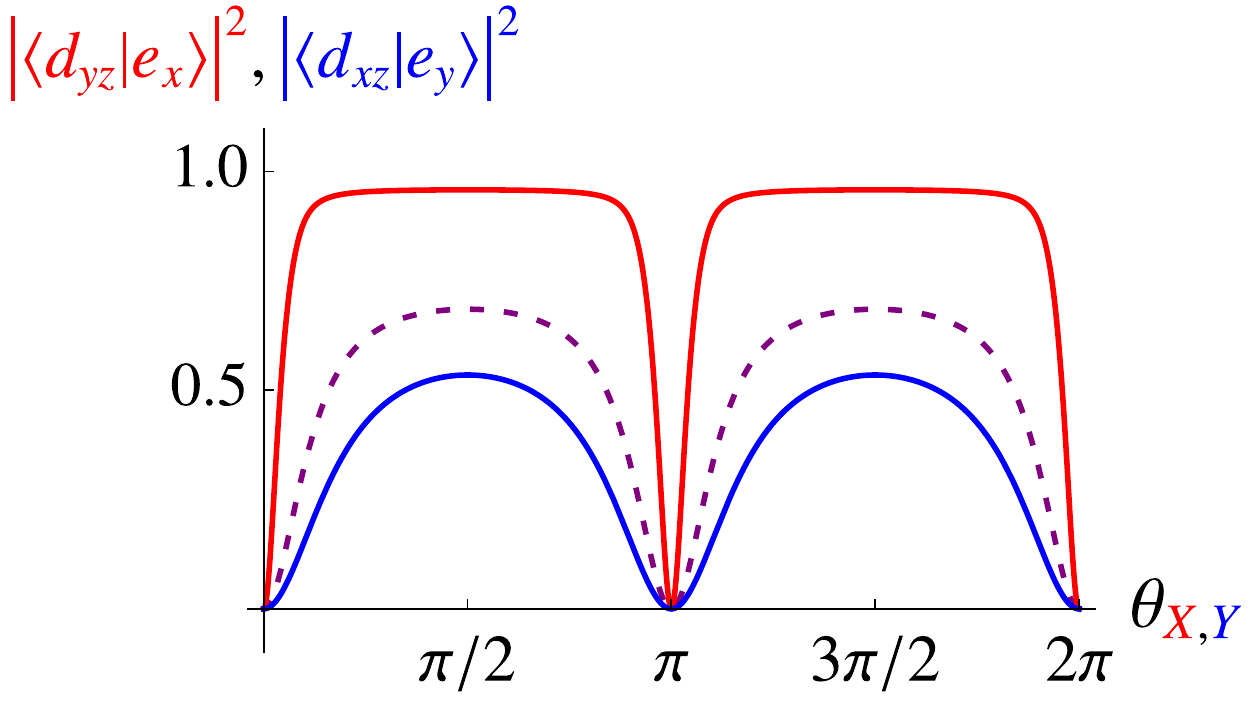}}
\caption{The change of orbital weight on the hole pocket
(a)
and on the electron pockets (b) between the tetragonal phase
(dashed line) and the nematic phase (solid lines) The angle $\theta_{X}$
($\theta_{Y}$) is measured with respect to the $X$ ($Y$)
direction.
For the hole pocket, we present the results including the SOC ${\bar \lambda}$ (the band splitting at  $k_{x,y}=0$ is $\pm \sqrt{{\bar \Phi}^2 + {\bar \lambda}^2/4}$).  We used ${\bar \lambda} =2 {\bar \Phi}_h$. }
\label{Fig:Orbital}
\end{figure}

  For the electron pockets
we find that
 the relations $\cos{\phi_{X,Y}}=A\sin{\theta_{X,Y}}$
also hold, but $A$ becomes different for $X$ and $Y$ pockets. We
define the dimensionless $\Phi_{e}$ via $A_{X}\approx A(1-\Phi_{e})$
and $A_{Y}\approx A(1+\Phi_{e})$, up to $O(\Phi_{e}^{2})$ terms.
To match ARPES and STM data for the shapes of the $h$ and $X$ pockets,
$\Phi_h$ must be positive and $\Phi_e$ negative.
 A  positive $\Phi_h$ increases
the $d_{xz}$ spectral weight on the hole pocket, particularly when $\Phi_h >1/2$, see Fig.~\ref{Fig:Orbital}a.
 At $\Phi_e \sim 1$ the hole pocket is almost entirely $d_{xz}$.
A negative $\Phi_{e}$ increases the weight of the $d_{yz}$ orbital on
the the $X$ pocket and reduces the weight of the $d_{xz}$ orbital on the
$Y$ pocket, as shown in Fig.~\ref{Fig:Orbital}b.
 We computed
the dimensionless $\Phi_{h,e}$ using ${\bar \Phi}_h =10 \mathrm{meV}$,  $|{\bar \Phi}_e| \sim 20 \mathrm{meV}$ (Refs. ~\cite{borisenko,Coldea16,BrianSTM17})
 and band structure parameters that
fit the ARPES data for the $Z$ pocket~\cite{Coldea16,ColdeaPrivate} in the nematic phase above $T_c$  and obtained~\cite{SM}
 $|\Phi_{e}|\sim 0.1, ~\Phi_{h}\sim 0.3$.
For such $\Phi_h$ the orbital weight along the $Z$ pocket still interpolates between $d_{xz}$ and $d_{yz}$ and does not depend strongly on the SOC.  To
 simplify our analysis we then neglect SOC in the
 solution of the gap equations.

\emph{Pairing interaction.} The pairing interaction has three components
\textendash{} one involves fermions near the hole pocket, another
involves fermions near the two electron pockets, and the third one
is
 the pair hopping between hole and electron pockets. At the bare level
all three interactions are comparable,
 but the pair-hopping term
gets enhanced once one integrates out fermions with high energies\cite{RG,ckf,ruiqi,scalapino,BrianSTM17}.
This enhancement
can be understood as an indication of the system's tendency to
increase magnetic fluctuations at momenta connecting the
 $\Gamma$/$Z$
and the $X,Y$ points, consistent with the neutron scattering data \cite{neutron1,neutron2,neutron3}.
 We
  therefore  consider
only the pair hopping term for the pairing problem. In the band basis,
 the pair-hopping pairing interaction
 has the form
\begin{align}
 & H_{\mathrm{pair}}=h_{k}^{\dagger}h_{-k}^{\dagger}\left[U_{s}\left(e_{X,p}e_{X,-p}\cos^{2}\phi_{X}+e_{Y,p}e_{Y,-p}\cos^{2}\phi_{Y}\right)\right.\nonumber \\
 & \left.+U_{d}\cos{2\phi_{h}}\left(e_{X,p}e_{X,-p}\cos^{2}\phi_{X}-e_{Y,p}e_{Y,-p}\cos^{2}\phi_{Y}\right)\right]\label{4}
\end{align}
where repeated momentum indices are implicitly summed and spin indices
are omitted. In the tetragonal phase, $\cos^{2}\phi_{X,Y}=A^{2}(1-\cos{2\theta_{X,Y}})/2$,
$\phi_{h}=\theta_{h}$, and the two terms in (\ref{4}) describe pairing
interactions in the $s-$wave and $d-$wave channels with couplings
$U_{s}$ and $U_{d}$, respectively.
The ratio $U_{s}/U_{d} = (U+J)/(U-J) >1$ already at the bare level,
where $U$ and $J$ are Hubbard and Hund's interactions, and further
 increases under RG~\cite{ckf}. Then
 the leading instability in the absence of nematicity is towards
$s^{+-}$ superconductivity.

In the presence of nematic order the situation changes because now
$\cos{2\phi_{h}} \approx\cos{2\theta_{h}}-\Phi_{h}$ and $A_{X}\neq A_{Y}$. As a result, the $U_{d}$ term in (\ref{4})
acquires extra terms which have an ``$s-$wave" angular
dependence and effectively renormalize the $U_{s}$ term, making this
interaction different for fermions near the $X$ and $Y$ pockets.
Substituting the forms of $\cos{2\phi_{h}},\cos{2\phi_{X}}$ and $\cos{2\phi_{Y}}$
into (\ref{4}) and restricting to first-order terms in $\Phi_{h}$
and $\Phi_{e}$, we obtain the pairing interaction in the form

\begin{equation}
H_{\mathrm{pair}}=\frac{A^{2}}{2}\sum_{j=X,Y}h_{k}^{\dagger}h_{-k}^{\dagger}(A_{j}+B_{j}\cos{2\theta_{h}})e_{j,p}e_{j,-p}\label{5}
\end{equation}
where
\begin{align}
A_{X,Y} & =\left(1-\cos{2\theta_{X,Y}}\right)\left[U_{s}\left(1\mp2\Phi_{e}\right)\mp U_{d}\Phi_{h}\right]\nonumber \\
B_{X,Y} & =\pm U_{d}\left(1-\cos{2\theta_{X,Y}}\right)\left(1\mp2\Phi_{e}\right)\label{6}
\end{align}

\emph{Gap equations.} We use Eqs. (\ref{5}) and (\ref{6}) to
 obtain
 the linearized gap equations. The gap on the hole pocket is parametrized
by
$\Delta(\theta_{h})=\Delta_{h}(1+\alpha\cos{2\theta_{h}})$,
 (we neglect $\cos{4\theta}$ term to simplify presentation).
  The computational
steps are rather conventional
\cite{SM}.
 To  linear order in $\Phi_{e,h}$,
\begin{equation}
\alpha\approx \frac{U_{s}U_{d}}{U_{s}^{2}-U_{d}^{2}/2}\left(4|\Phi_{e}| - \frac{U_{d}}{U_{s}}\Phi_{h}\right)\label{7}
\end{equation}
Notice that $\alpha$ depends only on the ratio $U_{d}/U_{s}$, and
not on the strength of the interaction, which is compensated by the
Cooper logarithm.

We see that there are two contributions to the gap anisotropy $\alpha$,
originating from the components of the nematic order on hole and electron
pockets. Because $\Phi_{h}$ and $\Phi_{e}$ have opposite signs,
the sign of $\alpha$ depends on
 their strength and on the ratio between the interactions $U_{d}/U_{s}$.
  Because $4 |\Phi_e| > \Phi_h$  and $U_d/U_s \leq 1$, we find   $\alpha \sim 0.2$ is positive, i.e., the gap $\Delta_{h}(\theta_{h})$ has its maximum
along the $X$ direction $\theta_{h}=0$. This is consistent
with the STM and ARPES data~\cite{Dong16,Shibauchi15,shin,BrianSTM17,Borisenko18,Zhou18,Rhodes18}.
The $\Phi_h$ term in (\ref{7}) is further reduced if we take into account the fact that
 the ratio $U_{s}/U_{d}$ grows under the renormalization group flow~\cite{ckf}.

To go beyond this analytic expansion in powers of $\Phi_{e,h}$, we
solved  the gap equations numerically for the same set of parameters, but
 not restricting to first order in $\Phi_{h,e}$.
  We found the same gap structure but somewhat larger $\alpha \approx0.65$.
 The
 result is shown in Fig.~\ref{Fig:HoleGap} along with the
STM data from Ref. \cite{BrianSTM17}.
For this plot, we
 added to $\Delta (\theta_h)$  additional $\beta \cos4\theta_{h}$
 term with $\beta =-0.1$.
  The $\cos4\theta_{h}$ dependence arises already in the tetragonal phase and is determined by details beyond our model.

The sign of the gap anisotropy can be interpreted as the indication
that in the nematic state the pairing interaction between the $h$ and
$X$ pockets becomes stronger than between the $h$ and $Y$ pockets.
Because the positive contribution to $\alpha$ comes
from $\Phi_{e}$, the
increase of the $h-X$ interaction can
be traced back to the increase of
$d_{yz}$ orbital weight on the $X$ pocket. In this respect, qualitatively our results
agree with Refs. \cite{BrianSTM17,Brian17}, where the increase of
the $d_{yz}$ orbital weight was introduced phenomenologically,
via an orbital dependent constant $Z$-factor.
However, in
our theory the modification of the $d_{xz}/d_{yz}$ orbital weights naturally
emerges
within the low-energy model and does not require
the inclusion of additional $Z-$factors.

On the electron pockets, to leading order in $\Phi_{h,e}$, the gaps
have the forms
$\Delta_{X,Y}=-\Delta_{h}\gamma_{X,Y}(1-\cos{2\theta_{X,Y}})$, where
$\gamma_{X,Y}=\gamma\left[1\pm (2|\Phi_{e}| -U_{d}/U_{s}\Phi_{h} +\alpha/2)\right]$
and $\gamma >0$
 is a number whose value depends on the electronic structure. The vanishing of
the gaps at $\cos{2\theta_{X,Y}}=\pm1$ is an artifact of neglecting
the $d_{xy}$ orbital in the pairing problem.
 In reality,
 the gaps $\Delta_{X,Y}$ tend
to small but finite values along the $X$ and $Y$ directions,
respectively. The ARPES and STM data
reported an anisotropic,
but still sign-preserving gap on the $X$ pocket, with gap maximum
at $\theta_{X}=\pi/2$, consistent with our formulas. The overall
sign of $\Delta_{X,Y}$ is opposite to that of $\Delta_{h}$. The
dependence of $\gamma_{X,Y}$ on the nematic order shows that the
gap magnitude is larger on the $X$ pocket than on the $Y$ pocket.
We propose to verify this in future experiments.

\emph{Fermionic self-energy.} The STM data
 indicate that in the
nematic phase the $Y$ pocket is less visible than the $X$ pocket, and
 in
some ARPES studies \cite{watson,Rhodes18} this $Y$ pocket
  has  not been observed. To understand
this feature, we computed the self-energy on both electron pockets
to second order in $U_{s}$ and $U_{d}$ and extracted the
 actual
 quasiparticle
residues $Z_{X,Y}$ on each electron pocket~\cite{SM}. We find $Z_{Y}>Z_{X}$
simply because the effective interaction is larger for fermions on
the $X$ pocket (we recall that larger interaction
 leads to a smaller $Z$). If this was the only effect, we would expect the
$Y$ pocket to become more visible. However, like we said, nematic order
also increases the $d_{yz}$ spectral weight of the $X$ pocket and
decreases the $d_{xz}$ orbital spectral weight of the $Y$ pocket
 (see Fig. \ref{Fig:Orbital}).
 If the $d_{xy}$ orbital excitations
are not observed in STM and ARPES because of matrix elements, or if
the $d_{xy}$ orbital is more incoherent than the $d_{xz}/d_{yz}$
orbitals ~\cite{kotliar,bascones,Medici,Fanfarillo17}, then the $Y$
pocket should indeed become less visible in the nematic phase. We caution, however, that recent  ARPES study~\cite{Rhodes18} did not find $d_{xy}$ excitations
 on the $X$ pockets to be more incoherent that $d_{yz}$ excitations, so the
 reason why the $Y$ pocket is less visible in STM and some ARPES studies is not yet understood.

\emph{Conclusions.} In this paper we argued that the experimentally
observed anisotropy of the superconducting gap in bulk FeSe can be
explained within the low-energy model for nematic order, without adding
phenomenologically different quasiparticle weights for the $d_{xz}/d_{yz}$
orbitals. Our key result is that $T_{c}$ is not strongly affected
by the nematic order, but nematicity mixes $s-$wave and $d-$wave pairing
channels and gives rise to a $\cos{2\theta_{h}}$ gap anisotropy on
the hole pocket
The sign of the $\cos{2\theta_{h}}$ term is determined by the interplay
between the nematic order parameters on hole and electron pockets,
which are of different sign, and the relative strength of $s-$wave
and $d-$wave components of the pairing interaction.
On the $Z$ pocket, we found a sizable $\cos{2\theta_{h}}$ gap anisotropy
 with the gap maximum along the $X$ direction,
in agreement with the data. In our calculations the gap on the $\Gamma$ pocket is smaller and less anisotropic.
 On the peanut-like $X$ pocket, the gap
is found to be maximal along the minor axis, which is also in agreement
with the data. We also argued that nematicity decreases the weight
of the $d_{xy}$ orbital on the $X$ pocket and increases it on the
$Y$ pocket. This may potentially explain why the $Y$ pocket is less visible
in STM and
in some ARPES data.

\begin{acknowledgments}
We are thankful to B.~Andersen, L. Bascones, L. Benfatto, S. Borisenko, A. Coldea, M. Eschrig, P.~Hirschfield,
A.~Kreisel, C.~Meingast, L. Rhodes, J.C.~Séamus Davis, O.~Vafek, M. Watson,
and Y.~Y.~Zhao for useful discussions. JK was supported by the National
High Magnetic Field Laboratory through NSF Grant No.~DMR-1157490
and the State of Florida. RMF and AVC were supported by the Office
of Basic Energy Sciences, U.S. Department of Energy, under awards
DE-SC0012336 (RMF) and DE-SC0014402 (AVC). J.K.~thanks FTPI at the
University of Minnesota for hospitality during the completion of this
work. The authors are thankful to KITP at UCSB, where part of the
work has been done. KITP is supported by NSF grant PHY 17-48958.
\end{acknowledgments}

\vspace{0.5cm}

\begin{widetext}
\begin{center}
\textbf{\large{}{}{}{}{}{}Supplementary material for ``Anisotropic
superconductivity in FeSe without orbital selectivity''}{\large{}{}{}{}{}{}
}
\par\end{center}
\setcounter{equation}{0} \setcounter{figure}{0} \setcounter{table}{0}

\global\long\def\theequation{S\arabic{equation}}

\global\long\def\thetable{S\arabic{table}}
 \global\long\def\thefigure{S\arabic{figure}}

\section{details of the low-energy model}

\subsection{Hole Pockets}

The dispersion near the hole pockets centered at $k_{x,y}$ is expressed
in terms of the two-component spinor $\psi_{\Gamma}=(d_{xz},\ d_{yz})^{T}$.
We follow Refs. \cite{Vafek13,review_fc} and write the Hamiltonian in the
tetragonal phase in the absence of spin-orbit coupling (SOC) as
\begin{equation}
H_{h}^{(0)}=\psi_{h}^{\dagger}\left(\left(\epsilon_{h}-\frac{\fvec k^{2}}{2m_{h}}\right)\tau_{0}-\frac{b}{2}\big(k_{x}^{2}-k_{y}^{2}\big)\tau_{3}-2ck_{x}k_{y}\tau_{1}\right)\psi_{h}=\psi_{h}^{\dagger}
\big(H_{0}\tau_{0}-\frac{b}{2}k^{2}\cos2\theta_{h}\tau_{3}-ck^{2}\sin2\theta_{h}\tau_{1}\big)\psi_{h}\ ,\label{EqnS:HoleBand}
\end{equation}
where $\theta_{h}$ is the angle measured with respect to the $X$
axis (we work in the 1-Fe Brillouin zone). The free parameters
of this Hamiltonian are shown in Table~\ref{TabS:HoleBand}, and are obtained from fitting
to ARPES data on the Z-pocket ($k_z =\pi$)~\cite{Coldea16,ColdeaPrivate}. All the energy parameters
are in units of meV, and the momentum are in units of the inverse lattice
constant.
\begin{table}[h]
\centering %
\begin{tabular}{| x{4em} | x{4em} | x{4em} | x{4em} | }
\hline
$\epsilon_{h}$  & $(2m_h)^{-1}$  & $b$  & $c$    \tabularnewline
\hline
13.6  & 473  & 529  & -265   \tabularnewline
\hline
\end{tabular}\caption{Band parameters of the hole pocket.}
\label{TabS:HoleBand}
\end{table}

The dispersion for the two hole pockets can easily be obtained numerically.  It is also instructive
to obtain an approximate analytical solution.
For this purpose, note that the parameters in the Table give $b\approx-2c>0$. In this case,
 the band dispersions around $Z$ can be approximated
as
\begin{equation}
H_{h}^{(0)}\approx H_{0}(k)\tau_{0}+H_{1}(k)\left(-\cos2\theta_{h}\tau_{3}+\sin2\theta_{h}\tau_{1}\right)
\end{equation}
with $H_{0}=\epsilon_{h}-k^{2}/(2m_{h})$ and $H_{1}(\fvec k)\approx bk^{2}/2$.
Diagonalization leads to Eq.~(1) of the main text with $\varphi_{h}=\theta_{h}$
and an isotropic dispersion, $\epsilon^Z_{1,2}=\epsilon_{h}- k^{2}/(2m_{h}) \pm bk^{2}/2$.   We will be mostly interested in the larger hole pocket.
Its dispersion is given by $\epsilon^Z_{1}=\epsilon_{h}- k^{2} (1/(2 m_{h}) - b/2)$.

In the nematic phase, the Hamiltonian acquires the extra term $H_{h^{(\mathrm{nem})}}={\bar \Phi}_{h} \psi_{h}^{\dagger}\tau_{3}\psi_h$
(see Ref. \cite{vafek_fernandes}). The total Hamiltonian $H_{h}=H_{h}^{(0)}+H_{h}^{(\mathrm{nem})}$
is then:
\begin{equation}
H_{h}\approx H_{0}(k)\tau_{0}+\left(\big({\bar \Phi}_{h}-H_{1}\cos2\theta_{h}\big)\tau_{3}+H_{1}\sin2\theta_{h}\tau_{1}\right)\quad\longrightarrow\quad H_{0}(k)\tau_{0}+H_{1}'\left(-\cos2\varphi_{h}\tau_{3}+\sin2\varphi_{h}\tau_{1}\right)
\end{equation}
with ($0 \leq \theta_h \leq \pi/2$)
\begin{equation}
\cot2\varphi_{h}\approx\frac{H_{1}\cos2\theta_{h}-{\bar \Phi}_{h}}{H_{1}\sin2\theta_{h}}=\cot2\theta_{h}-\frac{{\bar \Phi}_{h}}{H_{1}\sin2\theta_{h}}\quad\Longrightarrow\quad\varphi_{h}\approx\theta_{h}+\Phi_{h}\,\sin2\theta_{h}
\label{s1}
\end{equation}
where we defined the dimensionless nematic order parameter $\Phi_{h}\equiv {\bar \Phi}_{h}/2H_{1}\left(k_{F}\right)$, where
$H_1 \left(k_{F}\right) = b k^2_F/2$.
  Using the numbers from the Table,
we estimate on the $Z$ pocket
$2H_{1}\left(k_{F}\right)\approx30$ meV and ${\bar \Phi}_{h} \approx 10$meV. This yields  $\Phi_{h}\approx0.3$.

On the $\Gamma$ pocket, $k_F$ is smaller~\cite{Dong16,watson,Borisenko18,Rhodes18}, and for the same ${\bar \Phi}_{h}$, the dimensionless $\Phi_{h}$ is larger, at least by a factor of $2$.

In addition to the change in the orbital composition of the $\Gamma$
pocket, nematicity also deforms the shape of Fermi surface. The change
in the dispersion is $\delta\epsilon_{h}=-{\bar \Phi}_{h} \cos2\theta_{h}$,
giving rise to a change in the Fermi momentum $\delta k_{F}\sim-{\bar \Phi}_{h} \cos2\theta_{h}/v_{f}$.
Consequently, the Fermi pocket changes shape from circular to elliptical.
When ${\bar \Phi}_{h}$ is positive, its major axis points
along $Y$ direction, while the minor axis points along $X$ direction.

\subsubsection{Orbital content of the hole pocket}

In the tetragonal phase the weight of the $d_{xz}$ component along the larger hole
 pocket is $\sin^2{\theta_h}$ and the weight of $d_{yz}$ component is $\cos^2{\theta_h}$.
Along the $X$ direction, the orbital content is entirely $d_{yz}$, and along the $Y$ direction it is entirely $d_{xz}$.  In the nematic phase, the weight of $d_{xz}$ is
$\sin^2{\varphi_{h}}$ and the weight of $d_{yz}$ is
$\cos^2{\varphi_{h}}$.   Eq. (\ref{s1}) can be re-expressed as
\beq
\cot2\varphi_{h} \approx \frac{\cos2\theta_{h}-2 \Phi_{h}}{\sin2\theta_{h}}
\label{s2}
\eeq
An elementary analysis shows that the orbital weight at $\theta_h =0$ (i.e., along the $X$ direction) now depends on  whether $2\Phi_h <1$ or $2\Phi_h >1$. For smaller $\Phi_h$,
$\varphi_h (\theta_h=0) =0$, i.e., the weight along $X$ is entirely $d_{yz}$.  At $\theta_h = \pi/2$, $\varphi_h = \pi/2$, i.e., along $Y$, the orbital composition is $d_{xz}$, as in the absence of the nematicity.  At arbitrary $\theta_h$, $\varphi_h$ is different from $\theta_h$, and the orbital content changes compared to the one in the tetragonal phase.  We show the weight of $d_{xz}$ along the hole pocket in Fig. \ref{Fig:Orbital_1}a.  It increases in the nematic phase but still vanishes at $\theta_h =0, \pi$.

The situation changes when $\Phi_h >1/2$.  From Eq. (\ref{s2}) we now have $\varphi_h (\theta_h=0) =\pi/2$, i.e., the weight along $X$ is now entirely $d_{xz}$.  At $\theta_h = \pi/2$, we still have $\varphi_h = \pi/2$, i.e., the orbital weight is entirely $d_{xz}$.  This is a non-trivial change of orbital composition of the hole pocket in the tetragonal phase.  In Fig. \ref{Fig:Orbital_2_a} we show  the orbital content of $d_{xz}$ along the larger hole pocket at various $\Phi_h$. We see that along the $X$ direction it jumps from $0$ to $1$ between $\Phi_h <1/2$ and $\Phi_h >1/2$.

\begin{figure}[htbp]
\includegraphics[width=0.5\columnwidth]{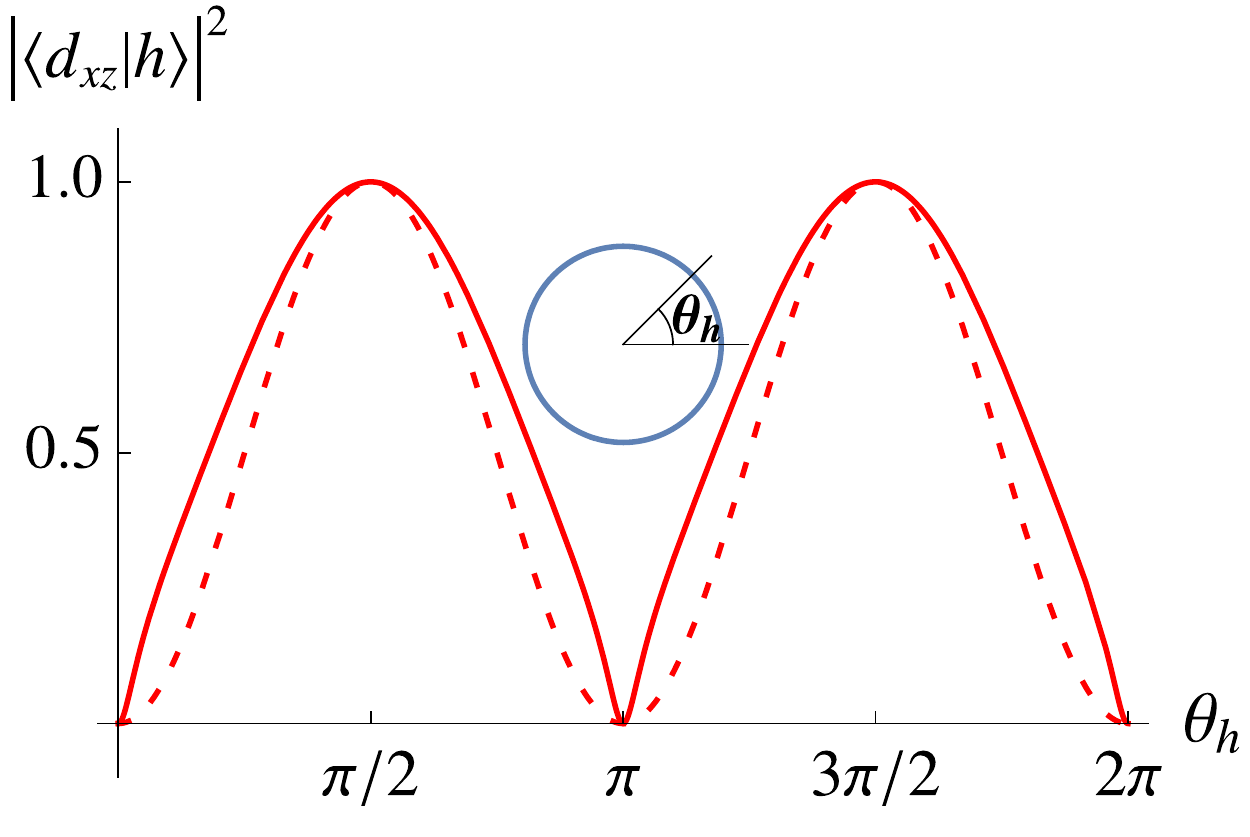}
\caption{The change of orbital weight on the hole pocket between the tetragonal phase
(dashed line) and the nematic phase (solid lines) in the absence of SOC.  The angle
 is measured with respect to the $X$
direction. }
\label{Fig:Orbital_1}
\end{figure}
\begin{figure}[htbp]
$\begin{array}{cc}
\includegraphics[width=0.48\columnwidth]{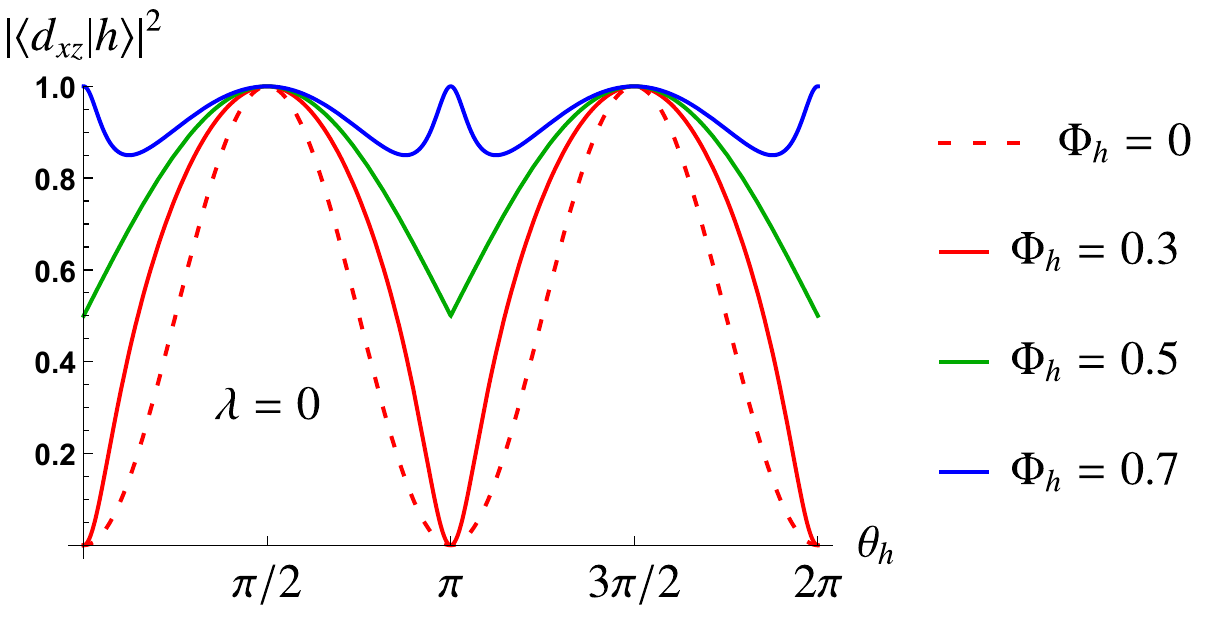}&
\includegraphics[width=0.48\columnwidth]{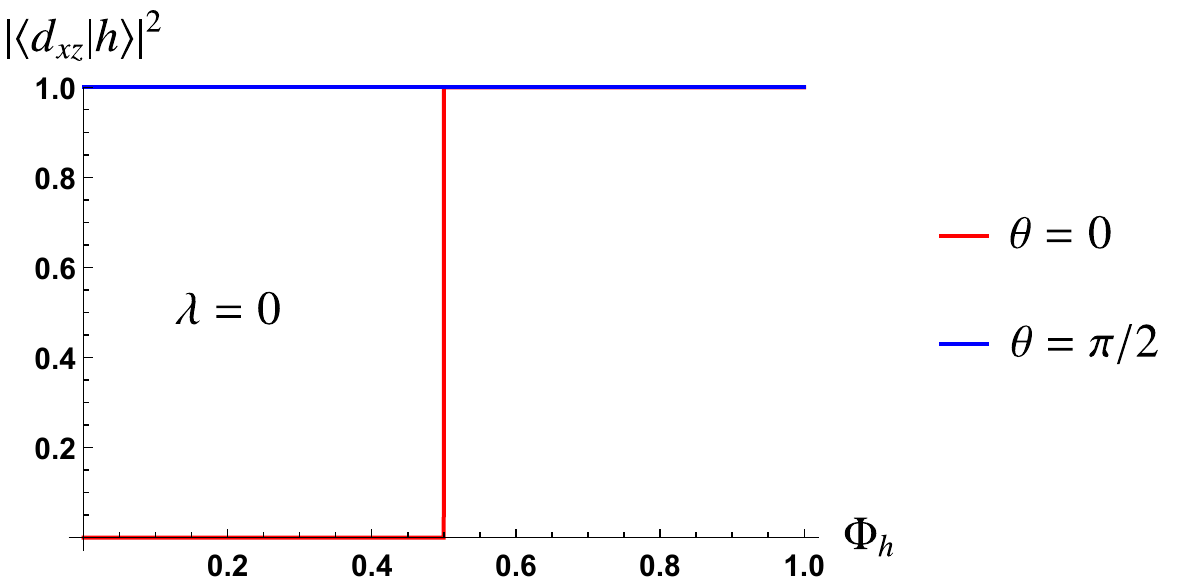}
\end{array}$
\caption{(a) The change of orbital weight on the hole pocket between the tetragonal phase
(dashed line) and the nematic phase (solid lines) for different $\Phi_h$ in the absence of SOC.    The angle is measured with respect to the $X$
direction. (b) The $d_{xz}$ orbital weight at $\theta_h =0$ and $\theta_h = \pi/2$ as a function of $\Phi_h$. }
\label{Fig:Orbital_2_a}
\end{figure}

We now add SOC. It gives rise to additional term in the quadratic form~\cite{Vafek13}
\beq
H_{h,SOC} =  \frac{\bar \lambda}{2} \psi_{h,\alpha}^{\dagger} \tau_2 \psi_{h,\beta} \sigma^3_{\alpha\beta}
\label{s3}
\eeq
 where $\alpha,\beta$ are spin indices, and $\tau_2$ acts on orbital indices. At  $k_{x,y}=0$, the splitting between the larger and smaller hole pockets is
now  $2 \sqrt{{\bar \Phi}^2_h + {\bar \lambda}^2/4}$.  One can easily check that in the presence of SOC the dispersions of the two hole pockets repel each other and do not cross along any direction (at $\lambda =0$ they necessary cross at some momentum).   Because of no-crossing, the smaller hole pocket sinks completely below the Fermi level when
$\sqrt{{\bar \Phi}^2_h + {\bar \lambda}^2/4} > \epsilon_h$.

Re-diagonalizing the quadratic form, we now obtain on a larger hole Fermi surface,  instead of (\ref{s2})
\beq
\cot2\varphi_{h} \approx \frac{\cos2\theta_{h}-2 \Phi_{h}}{\sqrt{\sin^2{2\theta_{h}} + \lambda^2}}
\label{s4}
\eeq
 where $\lambda = {\bar \lambda}/H_1 (k_F)$ is the dimensionless SOC constant.
 One can easily verify that now the orbital content along both $X$ and $Y$ directions is neither $d_{xz}$ nor $d_{yz}$, although along $Y$ it remains quite close to pure $d_{xz}$
  for realistic $\lambda \sim \Phi_h$.  We show the orbital weight of $d_{xz}$ along the hole pocket for several $\Phi_h$ and $\lambda = 2 \Phi_h$ in Fig. 3a of the main text. Here we show, in Fig.\ref{Fig:Orbital_3_a}  the evolution of the spectral weight of $d_{xz}$  with $\lambda$ for several $\Phi_h$.  In Fig. \ref{Fig:Orbital_4} we show the evolution of the $d_{xz}$ weight  at $\theta_h=0$ and $\theta_h = \pi/2$   as a function of $\Phi_h$ for $\lambda = 2 \Phi_h$.   Note the rapid increase of the spectral weight
   of $d_{xz}$ at $\theta_h=0$ around $\Phi_h =1/2$ and weak dependence on $\Phi_h$ of the $d_{xz}$ weight at $\theta_h = \pi/2$.  On the $Z$ pocket ($\Phi_h \sim 0.3$) the $d_{xz}$ weight at $\theta_h =0$ is rather small. However, if $\Phi_h$ on the $Z$ pocket is a bit larger, the weight increase towards $50\%$.
     At $\Phi_h \sim 0.7-0.8$, expected for the $\Gamma$ pocket, the weight of $d_{xz}$ at $\theta_h =0$ is around $80\%$. This agrees with the polarization ARPES analysis in~\cite{Rhodes18}.

\begin{figure}[htbp]
$\begin{array}{ccc}
\includegraphics[width=0.32\columnwidth]{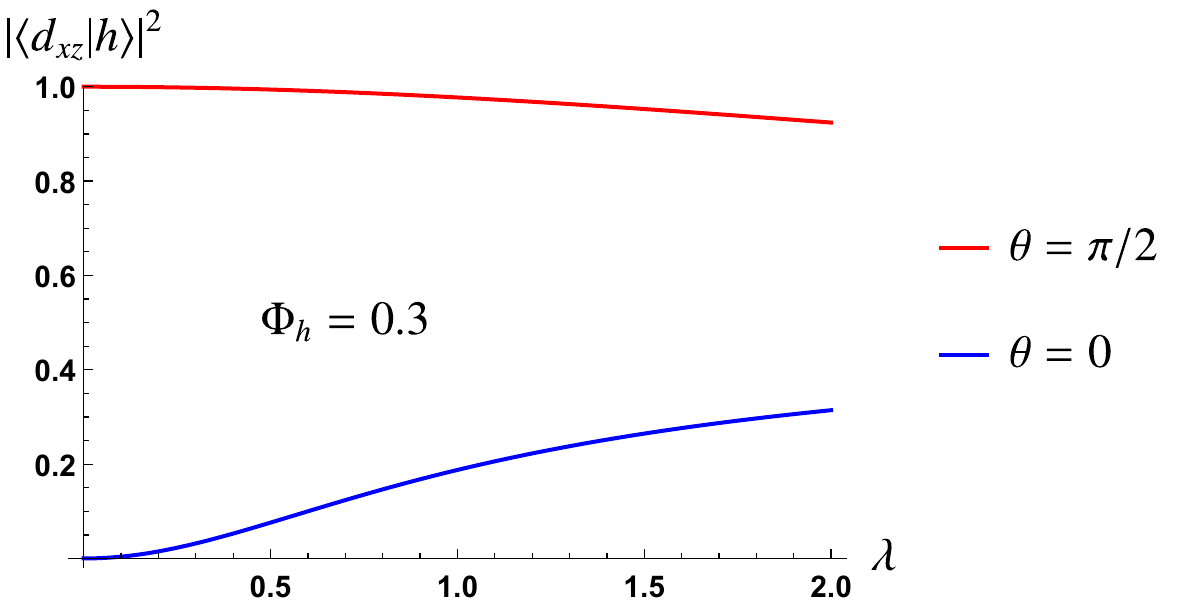}&
\includegraphics[width=0.32\columnwidth]{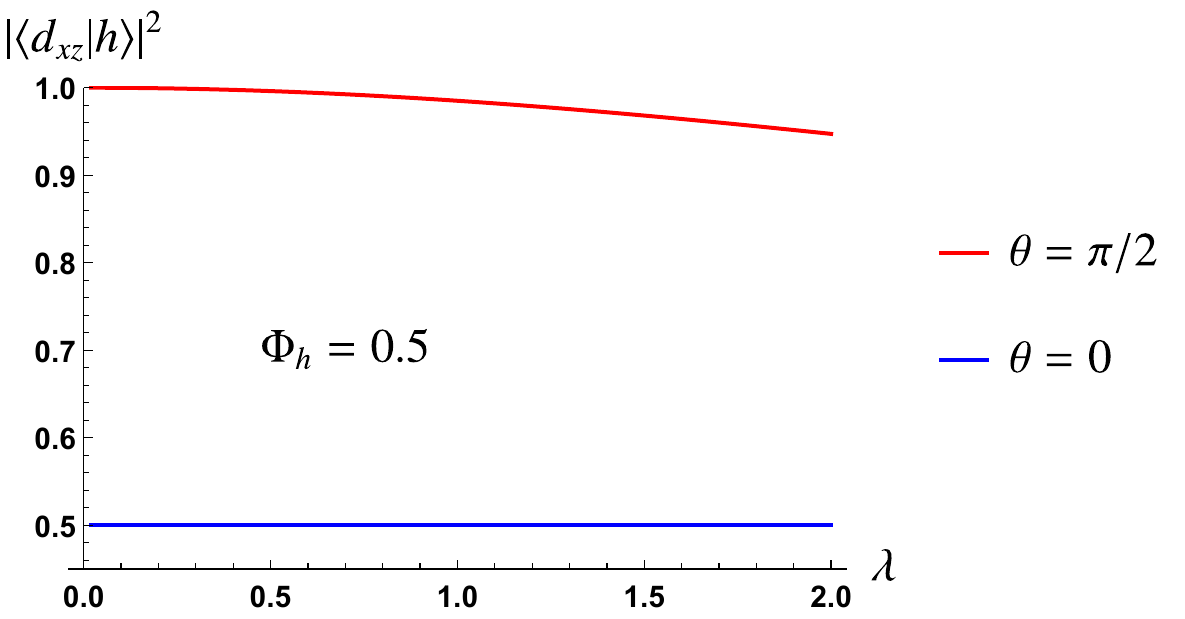}&
\includegraphics[width=0.32\columnwidth]{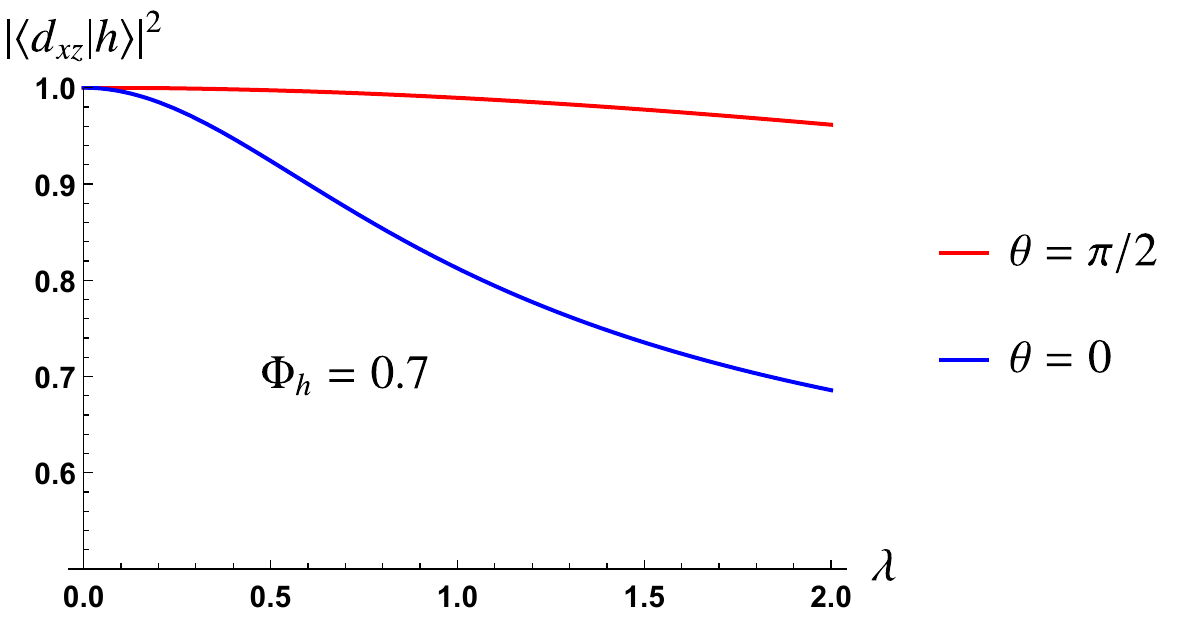}
\end{array}$
\caption{The evolution of the $d_{xz}$ orbital weight on the hole pocket with increasing $\lambda$ at three different $\Phi_h$ taken at and around critical $\Phi_h =1/2$.}
\label{Fig:Orbital_3_a}
\end{figure}
\begin{figure}[htbp]
\includegraphics[width=0.5\columnwidth]{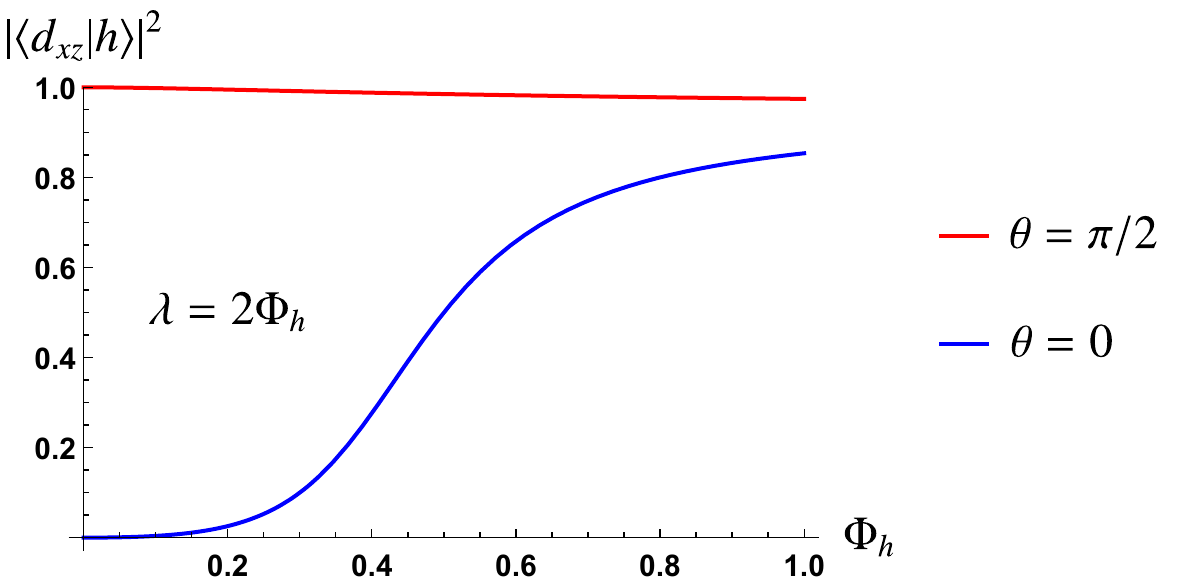}
\caption{The evolution of the $d_{xz}$ orbital weight on the hole pocket at $\theta_h=0$ and $\theta_h = \pi/2$  with increasing $\Phi_h$ and $\lambda =2\Phi_h$.}
\label{Fig:Orbital_4}
\end{figure}

\subsection{Electron Pockets}

For the electron pockets, an analytic expression similar to the case
of the hole pocket is not available. We start from the Hamiltonian~\cite{Vafek13,review_fc},
\begin{eqnarray}
H_{X,Y} & = & \Psi_{X,Y}^{\dagger}\begin{pmatrix}\epsilon_{1}+\frac{\fvec k^{2}}{2m_{1}}\mp\frac{a_{1}}{2}(k_{x}^{2}-k_{y}^{2}) & -iv_{X,Y}(\fvec k)\\
iv_{X,Y}(\fvec k) & \epsilon_{3}+\frac{\fvec k^{2}}{2m_{3}}\mp\frac{a_{3}}{2}(k_{x}^{2}-k_{y}^{2})
\end{pmatrix}\Psi_{X,Y}\\
v_{X}(\fvec k) & = & \sqrt{2}vk_{y}+\frac{p_{1}}{\sqrt{2}}\big(k_{y}^{3}+3k_{y}k_{x}^{2}\big)-\frac{p_{2}}{\sqrt{2}}k_{y}\big(k_{x}^{2}-k_{y}^{2}\big)\ ,\qquad v_{Y}(\fvec k)=\sqrt{2}vk_{x}+\frac{p_{1}}{\sqrt{2}}\big(k_{x}^{3}+3k_{x}k_{y}^{2}\big)+\frac{p_{2}}{\sqrt{2}}k_{x}\big(k_{x}^{2}-k_{y}^{2}\big)\nonumber
\end{eqnarray}
Here we list the band parameters fitted to ARPES experiments~\cite{Coldea16,ColdeaPrivate}.
All the energy parameters are in units of meV, and the momenta are
in units of the inverse lattice constant.
\begin{table}[htb]
\begin{centering}
\centering %
\begin{tabular}{ | x{3em} | x{3em} | x{4em} | x{4em} | x{3em} | x{3em} | x{3em} | x{3em} | x{3em} | }
\hline
$\epsilon_{1}$  & $\epsilon_{3}$  & $(2m_{1})^{-1}$  & $(2m_{3})^{-1}$  & $a_{1}$  & $a_{3}$  & $v$  & $p_{1}$  & $p_{2}$  \tabularnewline
\hline
-19.9  & -39.4  & 1.4  & 186  & 136  & -403  & -122  & -137  & -11.7  \tabularnewline
\hline
\end{tabular}
\par\end{centering}
\caption{Band parameters for the electron pockets.}
\label{TabS:EleBand}
\end{table}

Diagolizing the Hamiltonian numerically, we found that the
orbital composition of the $X$ electron pocket can be fitted using
the approximate form for the band operator in terms of the orbital
operators:
\begin{equation}
e_{X}=-iA\sin\theta_{X}d_{yz}+\sqrt{1-A^{2}\sin^{2}\theta_{X}}d_{xy}
\end{equation}

The value of $A$ can be estimated using the band parameters presented
below, yielding $A^{2}\approx0.7$. Note that there is another band
at the $X$ pocket that does not cross the Fermi level. The corresponding
operator, denoted here by $\tilde{e}_{X}$, is parametrized according
to:

\begin{equation}
\tilde{e}_{X}=i\sqrt{1-A^{2}\sin^{2}\theta_{X}}d_{yz}+A\sin\theta_{X}d_{xy}
\end{equation}

An important quantity for our analysis is the energy splitting $\Delta E$
between these two bands calculated at $k_{F}$ of the electron pocket.
Using the ARPES fitted parameters, we find $\Delta E\sim60$meV for
$\theta=\pi/2$.

Nematic order is included via:

\begin{equation}
H_{X,Y}^{\mathrm{(nem)}}=\mp\Delta_{e}^{(\mathrm{nem})}\Psi_{X,Y}^{\dagger}\begin{pmatrix}1 & 0\\
0 & 0
\end{pmatrix}\Psi_{X,Y}\ .
\end{equation}

To leading order of $\Delta_{e}^{(\mathrm{nem})}$, the wave function
of the upper band becomes
\begin{equation}
e_{X}'\approx e_{X}+\frac{\Delta_{e}^{(\mathrm{nem})}}{\Delta E}A\sin\theta_{X}\sqrt{1-A^{2}\sin^{2}\theta_{X}}\left(i\sqrt{1-A^{2}\sin^{2}\theta_{X}}d_{yz}+A\sin\theta_{X}d_{xy}\right)
\end{equation}

At $\theta_{X}=\pi/2$, where the spectral weight of $d_{yz}$ orbital
is maximum, we find that $e_{X}'$ can be expressed in the same form
of $e_{X}$ but with $A\rightarrow A_{X}=A+\delta A$, where:
\begin{equation}
\frac{\delta A}{A}\approx-\frac{\Delta_{e}^{(\mathrm{nem})}}{\Delta E}\big(1-A^{2}\big)\equiv-\Phi_{e}
\end{equation}

Here, we defined the dimensionless nematic order parameter $\Phi_{e}$.
Using the values of $A$ and $\Delta E$ mentioned above, and $\Delta_{e}^{(\mathrm{nem})}\approx-18$meV,
as indicated by ARPES measurements~\cite{amalia,borisenko}, we find
$\Phi_{e}\approx-0.1$.

\section{Pairing Interaction}

As explained in the main text, the RG analysis allows us to focus
only on two types of inter-pocket pairing interaction: the intra-orbital
pairing $U$ and inter-orbital pairing $J$. We find
\begin{eqnarray}
H_{\mathrm{pair}}^{(X)} & = & 2\sum_{\fvec k,\fvec p} h_{\fvec k\uparrow}^{\dagger} h_{\fvec-k\downarrow}^{\dagger} \big(1- t \cos4\phi_{h}\big)\big(U\cos^{2}\phi_{h}+J\sin^{2}\phi_{h}\big)e_{X-\fvec p\downarrow}e_{X\fvec p\uparrow}\cos^{2}\phi_{X}+h.c.\\
H_{\mathrm{pair}}^{(Y)} & = & 2\sum_{\fvec k,\fvec p}h_{\fvec k\uparrow}^{\dagger} h_{\fvec-k\downarrow}^{\dagger} \big(1- t \cos4\phi_{h}\big)\big(U\sin^{2}\phi_{h}+J\cos^{2}\phi_{h}\big)e_{Y-\fvec p\downarrow}e_{Y\fvec p\uparrow}\cos^{2}\phi_{Y}+h.c.
\end{eqnarray}
Instead of deriving how the angular dependence of the hole gap arises
due to SOC or renormalization, we introduce a phenomenological parameter
$-t\cos4\theta_{h}$ in the pairing interaction to account for
the angular dependence of the SC gap on $h$ even in the tetragonal
phase. Note that the $C_{4}$ symmetry is still conserved in the presence
of this term. In our numerical calculation, we set $t = 0.2 \ll 1$.

Adding them together yields:
\begin{eqnarray}
H_{\mathrm{pair}} & = & \sum_{\fvec k,\fvec p}h_{\fvec k\uparrow}^{\dagger}h_{-\fvec k\downarrow}^{\dagger}\big(1- t \cos4\phi_{h}\big)\left[U_{s}\left(e_{X,-\fvec p\downarrow}e_{X,\fvec p\uparrow}\cos^{2}\phi_{X}+e_{Y,-\fvec p\downarrow}e_{Y,\fvec p\uparrow}\cos^{2}\phi_{Y}\right)\right.\nonumber \\
 &  & \left.+U_{d}\cos{2\phi_{h}}\left(e_{X,-\fvec p\downarrow}e_{X,\fvec p\uparrow}\cos^{2}\phi_{X}-e_{Y,-\fvec p\downarrow}e_{Y,\fvec p\uparrow}\cos^{2}\phi_{Y}\right)\right]+h.c\label{EqnS:PairNem}
\end{eqnarray}
with $U_{s}=U+J$ and $U_{d}=U-J$. In our calculation, we set $U_{d}=U_{s}$.
Note that $\cos^{2}\phi_{X,Y}$ is the orbital weight of $d_{yz,xz}$
on the $e_{X,Y}$ band and $\cos^{2}\phi_{h}$ ($\sin^{2}\phi_{h}$)
are the weights of $d_{yz}$ ($d_{xz}$) orbitals on the hole band.
These weights can be obtained by diagonization of the matrix $H_{\Gamma,X,Y}$
with the nematic terms.

As shown in the previous sections, $\cos^{2}\phi_{X,Y}\approx A_{X,Y}^{2}\sin^{2}\theta_{X,Y}$
with $A_{X,Y}\approx A\big(1\mp\Phi_{e}\big)$ if only the first order
of the nematic order parameter $\Phi_{h,e}$ is kept in the expansion.
Additionally, $\cos2\phi_{h}\approx\cos2\theta_{h}+\Phi_{h}\cos4\theta_{h}-\Phi_{h}$.
For small nematicity and $t \ll1$, the $\cos4\theta_{h}$ terms
can be neglected in the pairing interaction. This gives
\begin{align}
H_{\mathrm{pair}}= & \frac{A^{2}}{2}\sum_{j=X,Y}h_{k}^{\dagger}h_{-k}^{\dagger}(1-\cos2\theta_{j})\big(A_{j}+B_{j}\cos{2\theta_{h}}\big)e_{j,p}e_{j,-p}\label{EqnS:ApproxPairNem}\\
\mbox{with}\quad A_{X,Y}= & \left[U_{s}(1\mp2\Phi_{e})\mp U_{d}\Phi_{h}\right]\ ,\ \mbox{and}\quad B_{X,Y}=\left[\pm U_{d}-2\Phi_{e}U_{d}\right]\label{EqnS:ApproxPairInt}
\end{align}

\section{Solution of the gap equations}

Starting with the pairing interaction presented in Eq.~\ref{EqnS:PairNem},
the linearized gap equation becomes (see Fig. \ref{FigS:SCSum})
\begin{align}
\Delta_{h}(\fvec k_{h})= & -(1-\beta\cos4\theta_{h})\left[(U_{s}+U_{d}\cos2\phi_{h})\int\frac{\rmd^{2}k}{(2\pi)^{2}}\frac{\tanh\big(\beta\epsilon_{X,\fvec k}/2\big)}{2\epsilon_{X,\fvec k}}\cos^{2}\phi_{X}\Delta_{X}\right.\nonumber \\
 & \quad\left.+(U_{s}-U_{d}\cos2\phi_{h})\int\frac{\rmd^{2}k}{(2\pi)^{2}}\frac{\tanh\big(\beta\epsilon_{Y,\fvec k}/2\big)}{2\epsilon_{Y,\fvec k}}\cos^{2}\phi_{Y}\Delta_{Y}\right]\label{EqnS:HoleGap}\\
\Delta_{X}(\fvec k_{X})= & -\cos^{2}\phi_{X}\int\frac{\rmd^{2}k}{(2\pi)^{2}}\frac{\tanh\big(\beta\epsilon_{h,\fvec k}/2\big)}{2\epsilon_{h,\fvec k}}(1-\beta\cos4\phi_{h})(U_{s}+U_{d}\cos2\phi_{h})\Delta_{h,\fvec k}\label{EqnS:EXGap}\\
\Delta_{Y}(\fvec k_{Y})= & -\cos^{2}\phi_{Y}\int\frac{\rmd^{2}k}{(2\pi)^{2}}\frac{\tanh\big(\beta\epsilon_{h,\fvec k}/2\big)}{2\epsilon_{h,\fvec k}}(1-\beta\cos4\phi_{h})(U_{s}-U_{d}\cos2\phi_{h})\Delta_{h,\fvec k}\label{EqnS:EYGap}
\end{align}
We numerically solved this equation using the band structure and interaction
parameters discussed above and the results are presented in Fig.~\ref{Fig:HoleGap}. We found that the gap on the hole pocket is roughly proportional to $1 + 0.65 \cos2\theta_h - 0.1 \cos4\theta_h$.

\begin{figure}[htbp]
\raisebox{0.5cm}{\includegraphics[scale=0.7]{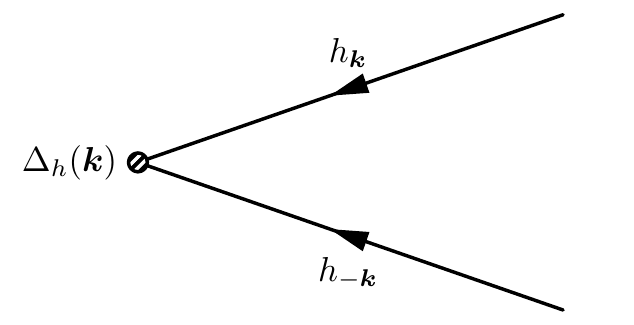}} \raisebox{1.5cm}{\LARGE{}$=$}
\includegraphics[scale=0.9]{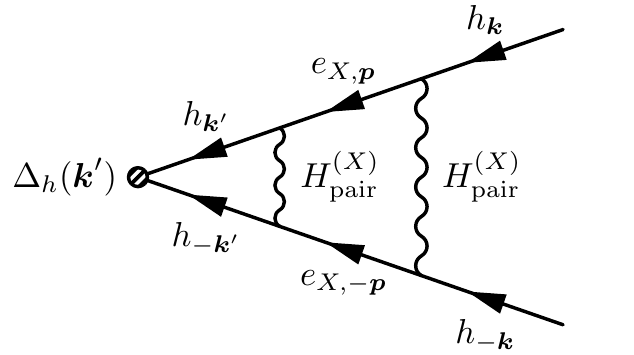} \raisebox{1.5cm}{$+$}
\includegraphics[scale=0.9]{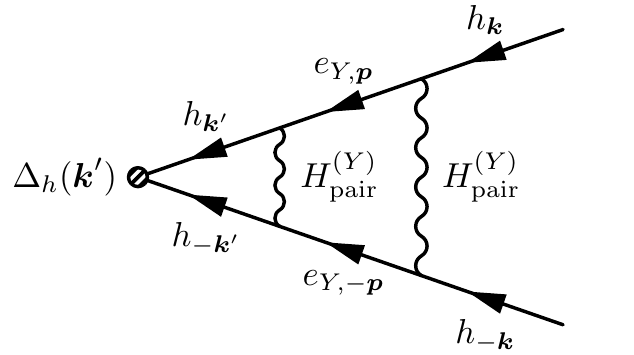} \caption{The gap equation for the hole pocket at $T=T_{c}$, with spin indices suppressed.}
\label{FigS:SCSum}
\end{figure}

We proceed now with the derivation of the analytical expression of
the SC gap in leading order in $\Phi_{h,e}$. For this purpose, we
start with the approximated pairing interaction in Eq.~\ref{EqnS:ApproxPairNem}.
The pairing equations become
\begin{align}
\Delta_{h}(\fvec k_{h})= & -A^{2}\left[(A_{X}+B_{X}\cos2\theta_{h})\int\frac{\rmd^{2}k}{(2\pi)^{2}}\frac{\tanh\big(\beta\epsilon_{X,\fvec k}/2\big)}{2\epsilon_{X,\fvec k}}\sin^{2}\theta_{X}\Delta_{X}\right.\nonumber \\
 & \quad\left.+(A_{Y}+B_{Y}\cos2\theta_{h})\int\frac{\rmd^{2}k}{(2\pi)^{2}}\frac{\tanh\big(\beta\epsilon_{Y,\fvec k}/2\big)}{2\epsilon_{Y,\fvec k}}\sin^{2}\theta_{Y}\Delta_{Y}\right]\nonumber \\
\Delta_{X}(\fvec k_{X})= & -A^{2}\sin^{2}\theta_{X}\int\frac{\rmd^{2}k}{(2\pi)^{2}}\frac{\tanh\big(\beta\epsilon_{h,\fvec k}/2\big)}{2\epsilon_{h,\fvec k}}(A_{X}+B_{X}\cos2\theta_{h})\Delta_{h,\fvec k}\nonumber \\
\Delta_{Y}(\fvec k_{Y})= & -A^{2}\sin^{2}\theta_{Y}\int\frac{\rmd^{2}k}{(2\pi)^{2}}\frac{\tanh\big(\beta\epsilon_{h,\fvec k}/2\big)}{2\epsilon_{h,\fvec k}}(A_{Y}+B_{Y}\cos2\theta_{h})\Delta_{h,\fvec k}
\end{align}

We can then parametrize the gaps as
\begin{equation}
\Delta_{h}=\Delta_{1}+\Delta_{2}\cos2\theta_{h}\ ,\quad\Delta_{X}=\Delta_{3}\sin^{2}\theta_{X}\,\mbox{and}\quad\Delta_{Y}=\Delta_{4}\sin^{2}\theta_{Y}\ .
\end{equation}

The gap equations for $\Delta_{i}$ can be written in a matrix form.
To further simplify the notation, we define
\begin{align}
\Xi_{X}= & \int\frac{\rmd^{2}k}{(2\pi)^{2}}\frac{\tanh\big(\beta\epsilon_{X,\fvec k}/2\big)}{2\epsilon_{X,\fvec k}}\sin^{4}\theta_{X} & \Xi_{Y}= & \int\frac{\rmd^{2}k}{(2\pi)^{2}}\frac{\tanh\big(\beta\epsilon_{Y,\fvec k}/2\big)}{2\epsilon_{Y,\fvec k}}\sin^{4}\theta_{Y}\nonumber \\
\Xi_{h,j}= & \int\frac{\rmd^{2}k}{(2\pi)^{2}}\frac{\tanh\big(\beta\epsilon_{h,\fvec k}/2\big)}{2\epsilon_{h,\fvec k}}\big(\cos2\theta_{h}\big)^{j}\quad\mbox{with}\ j=0,1,2\ .
\end{align}

All these integrals are $O(\ln(\Lambda/T))$, which is given by the
band dispersion in the tetragonal phase. In the weak coupling limit,
we therefore can keep only the logarithmic term to expand the SC gap
to the leading order of nematicity and neglect the change of band
dispersion by the nematicity. Therefore, $\epsilon_{e_{X}}(\theta)=\epsilon_{e_{Y}}$
and $\epsilon_{h}$ is still $C_{4}$ symmetric, leading to $\Xi_{X}=\Xi_{Y}=\Xi_{e}$
and $\Xi_{h,1}=0$. Furthermore, with quadratic hole dispersion, $\Xi_{h,2}=\Xi_{h,0}/2$.
$\Xi_{h,0}=\Pi_{h}$ is the usual particle-particle bubble for the
hole pocket. The matrix equations for $\Delta_{j}$ becomes
\begin{equation}
\begin{pmatrix}\Delta_{1}\\
\Delta_{2}
\end{pmatrix}=-A^{2}\Xi_{e}\begin{pmatrix}A_{X} & A_{Y}\\
B_{X} & B_{Y}
\end{pmatrix}\begin{pmatrix}\Delta_{3}\\
\Delta_{4}
\end{pmatrix}\ ,\qquad\begin{pmatrix}\Delta_{3}\\
\Delta_{4}
\end{pmatrix}=-A^{2}\Pi_{h}\begin{pmatrix}A_{X} & B_{X}/2\\
A_{Y} & B_{Y}/2
\end{pmatrix}\begin{pmatrix}\Delta_{1}\\
\Delta_{2}
\end{pmatrix}
\end{equation}

To solve this equation, we write
\begin{equation}
\begin{pmatrix}\Delta_{1}\\
\Delta_{2}
\end{pmatrix}=A^{4}\Xi_{e}\Pi_{h}\begin{pmatrix}A_{X} & A_{Y}\\
B_{X} & B_{Y}
\end{pmatrix}\begin{pmatrix}A_{X} & B_{X}/2\\
A_{Y} & B_{Y}/2
\end{pmatrix}\begin{pmatrix}\Delta_{1}\\
\Delta_{2}
\end{pmatrix}=\big(M_{0}+M_{1}\big)\begin{pmatrix}\Delta_{1}\\
\Delta_{2}
\end{pmatrix}\ ,
\end{equation}
where the matrix $M_{0}$ contains no nematic terms, and $M_{1}\sim O(\Phi_{h,e})$.
With the expression of $A_{X,Y}$ and $B_{X,Y}$ in Eq.~\ref{EqnS:ApproxPairInt},
we find
\begin{equation}
M_{0}=2A^{4}\Xi_{e}\Pi_{h}\begin{pmatrix}U_{s}^{2} & 0\\
0 & U_{d}^{2}/2
\end{pmatrix}\ ,\qquad M_{1}=-A^{4}\Xi_{e}\Pi_{h}\big(2U_{d}^{2}\Phi_{h}+8U_{s}U_{d}\Phi_{e})\begin{pmatrix}0 & 1/2\\
1 & 0
\end{pmatrix}
\end{equation}

We see $\Delta_{1}$ and $\Delta_{2}$ decouples in $M_0$, reflecting that
$s-$ and $d-$wave are two pairing instabilities. When $U_{s}>U_{d}$,
$s$-wave is the leading solution, with $T_{c}$ given by the equation
$2A^{4}U_{s}^{2}\Xi_{e}\Pi_{h}=1$. $s-$wave and $d-$wave are mixed
due to the perturbative term $M_{1}$. To leading order in $M_{1}$,
the solution is given by
\begin{equation}
\alpha=\frac{\Delta_{2}}{\Delta_{1}}\approx-\frac{U_{s}U_{d}}{U_{s}^{2}-U_{d}^{2}/2}\left(4\Phi_{e}+\frac{U_{d}}{U_{s}}\Phi_{h}\right)\ .
\end{equation}
As we said before, we set $\Phi_{e}\sim-0.1$, $\Phi_{h}=0.3$,
and $U_{d}/U_{s} \leq 1$. This yields $\alpha\approx0.2$.  We recall that we used the dispersion for the Z-pocket ($k_z = \pi$).
 If we don't expand the gap to leading order in $\Phi_h$ and $\Phi_e$ and  instead
 solve  Eqs.~\ref{EqnS:HoleGap}\textendash \ref{EqnS:EYGap} numerically, we obtain larger $\lambda \sim 0.65$, which we cited in the main text.  We present the result of  the numerical solution in Fig. 2 of the main text. For completeness, we also present the result at the $\Gamma$ point. Here $\Phi_h$ is larger and, hence, the gap anisotropy is smaller. We also found that the gap magnitude is reduced if we use the same values of $U_s$ and $U_d$ at $\Gamma$ and at $Z$. We show the results in Fig.~\ref{Fig:Orbital_5}. A smaller gap at $\Gamma$ is consistent with ARPES results in Refs. \cite{Dong16,Borisenko18}. The authors of \cite{Rhodes18}, however,  reported the gap at $\Gamma$ larger than that at $Z$.

 \begin{figure}[htbp]
\includegraphics[width=0.6\columnwidth]{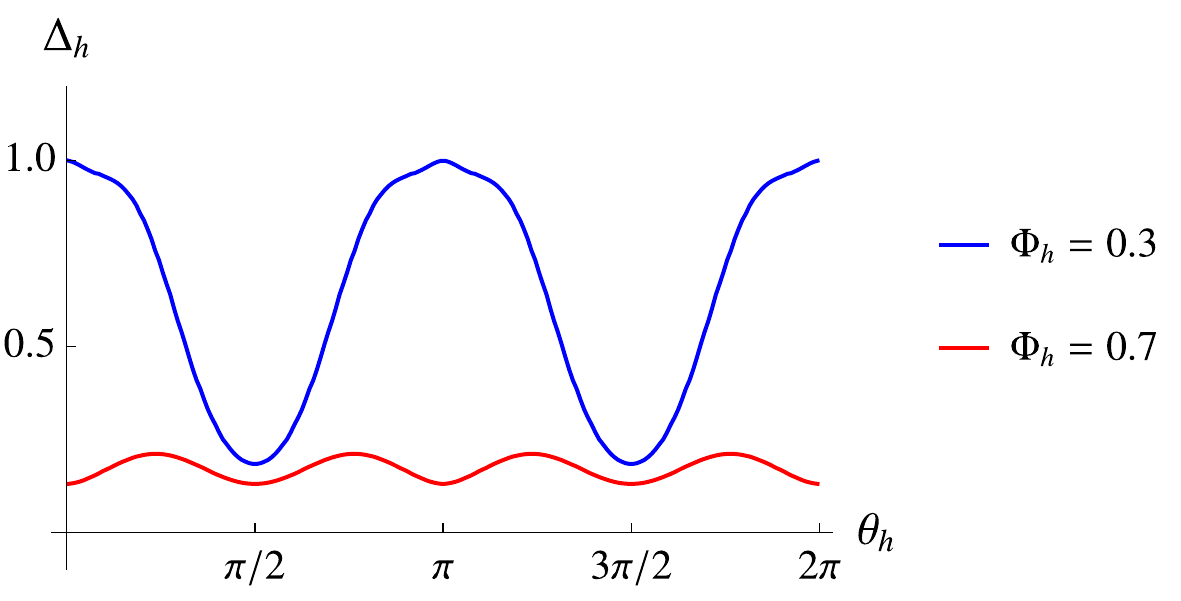}
\caption{Superconducting gap on the $Z$ pocket, in comparison with that on the $\Gamma$ pocket (red). We used the same pairing interactions $U_s$ and $U_d$ at $\Gamma$ and $Z$, same $\Phi_e$ and larger $\Phi_h$ at $\Gamma$ than at $Z$.}
\label{Fig:Orbital_5}
\end{figure}

For the electron pockets, we find:
\begin{align}
\Delta_{e_{X}}= & -A^{2}\sin^{2}\theta_{X}\Pi_{h}\left(A_{X}\Delta_{1}+\frac{B_{X}}{2}\Delta_{2}\right)\approx-A^{2}\Pi_{h}\sin^{2}\theta_{X}U_{s}\Delta_{1}\left[1-2\Phi_{e}-\frac{U_{d}}{U_{s}}\left(\Phi_{h}-\frac{\alpha}{2}\right)\right]\\
\Delta_{e_{Y}}= & -A^{2}\sin^{2}\theta_{Y}\Pi_{h}\left(A_{Y}\Delta_{1}+\frac{B_{Y}}{2}\Delta_{2}\right)\approx-A^{2}\Pi_{h}\sin^{2}\theta_{Y}U_{s}\Delta_{1}\left[1+2\Phi_{e}+\frac{U_{d}}{U_{s}}\left(\Phi_{h}-\frac{\alpha}{2}\right)\right]
\end{align}
The vanishing of the gap at the edges of the pockets is the artefact of neglecting $d_{xy}$ orbital.  Once it it included, the gap remains highly anisotropic, but does not vanish anywhere on $X$ ($Y$)  pocket.
\end{widetext}


\begin{thebibliography}{10}
\bibitem{comm_a}  We did not set the pairing interaction to be different for $d_{xz}$ and $d_{yz}$ orbitals due to nematic order. Such renormalization is rather weak if one introduces orbital order and directly compute the splitting of interactions on $d_{xz}$ and $d_{yz}$ orbitals
    ~\cite{BrianSTM17,Fanfarillo17}. The situation may be different if nematicity is due to composite Ising spin order~\cite{Benfatto18}.
\end{thebibliography}
\end{document}